\documentclass[sigconf]{acmart}
\copyrightyear{2024}
\acmYear{2024}
\setcopyright{acmlicensed}\acmConference[ASE '24]{39th IEEE/ACM International Conference on Automated Software Engineering }{October 27-November 1, 2024}{Sacramento, CA, USA}
\acmBooktitle{39th IEEE/ACM International Conference on Automated Software Engineering (ASE '24), October 27-November 1, 2024, Sacramento, CA, USA}
\acmDOI{10.1145/3691620.3695271}
\acmISBN{979-8-4007-1248-7/24/10}

\usepackage{enumitem}
\usepackage{pifont}
\usepackage{amsmath}
\usepackage{listings}
\usepackage{color,xcolor}
\usepackage{enumitem}
\usepackage[linesnumbered,ruled,vlined]{algorithm2e}
\usepackage{listings}
\usepackage{multirow}
\usepackage{tabularx}
\usepackage{subcaption}
\usepackage{wasysym}
\usepackage{makecell}
\usepackage{tcolorbox}

\AtEndPreamble{
	\usepackage{hyperref}

	\hypersetup{
		colorlinks = true,
		linkcolor = purple,
		anchorcolor = purple,
		citecolor = purple,
		filecolor = purple,
		urlcolor = purple
	}
}

\newcommand{\find}[1]{
\begin{tcolorbox}[leftrule=0.5mm,toprule=0mm,bottomrule=0mm,left=0.7pt,right=0.7pt,top=0.2pt,bottom=0.2pt]
\em #1
\end{tcolorbox}
}

\newcommand{\toolname}{\textsc{MalHug}\xspace}

\definecolor{codegreen}{HTML}{2ba02b}
\definecolor{codered}{HTML}{ff1b1c}
\definecolor{redbrush}{RGB}{192, 0, 0}
\definecolor{greenbrush}{RGB}{113, 173, 71}
\newcommand{\cmark}{\textcolor{greenbrush}{\ding{51}}}
\newcommand{\xmark}{\textcolor{redbrush}{\ding{55}}}

\begin{document}

\title{Models Are Codes: Towards Measuring Malicious Code Poisoning Attacks on Pre-trained Model Hubs}

% \author{Jian Zhao\textsuperscript{1}\textsuperscript{$\ast$}, Shenao Wang\textsuperscript{1}\textsuperscript{$\ast$}, Yanjie Zhao\textsuperscript{1}\textsuperscript{$\dag$}, Xinyi Hou\textsuperscript{1}, Kailong Wang\textsuperscript{1}, \\Peiming Gao\textsuperscript{2}, Yuanchao Zhang\textsuperscript{2}, Chen Wei\textsuperscript{2}\textsuperscript{$\dag$}, Haoyu Wang\textsuperscript{1}}

% \affiliation{%
%   \institution{\textsuperscript{1}~Huazhong University of Science and Technology}
%   \country{}
% }
% \email{{jian_zhao_,shenaowang,yanjie_zhao,xinyihou,wangkl,haoyuwang}@hust.edu.cn}
% \affiliation{%
%   \institution{\textsuperscript{2}~Mybank, Ant Group}
%   \country{}
% }
% \email{{peiming.gpm,yuanchao.zhang,juyi.wc}@mybank.cn}

% \thanks{\textsuperscript{$\ast$}Both authors contributed equally to this research}
% \thanks{\textsuperscript{$\dag$}Yanjie Zhao~(yanjie\_zhao@hust.edu.cn) and Chen Wei~(juyi.wc@mybank.cn) are the corresponding authors.}

\author[J Zhao]{Jian Zhao}
\authornote{Both authors contributed equally to this research.}
\authornote{Hubei Key Laboratory of Distributed System Security, Hubei Engineering Research Center on Big Data Security, School of Cyber Science and Engineering, Huazhong University of Science and Technology.}
\email{jian_zhao_@hust.edu.cn}
\affiliation{%
  \institution{Huazhong University of Science and Technology}
  \city{Wuhan}           
  \country{China}
}

\author[S Wang]{Shenao Wang}
\authornotemark[1]
\authornotemark[2]
\email{shenaowang@hust.edu.cn}
\affiliation{%
  \institution{Huazhong University of Science and Technology}
  \city{Wuhan}           
  \country{China}
}

\author[Y Zhao]{Yanjie Zhao}
\authornote{Yanjie Zhao~(yanjie\_zhao@hust.edu.cn) and Chen Wei~(juyi.wc@mybank.cn) are the corresponding authors.}
\authornotemark[2]
\email{yanjie_zhao@hust.edu.cn}
\affiliation{%
  \institution{Huazhong University of Science and Technology}
  \city{Wuhan}           
  \country{China}
}

\author[X Hou]{Xinyi Hou}
\authornotemark[2]
\email{xinyihou@hust.edu.cn}
\affiliation{%
  \institution{Huazhong University of Science and Technology}
  \city{Wuhan}           
  \country{China}
}

\author[K Wang]{Kailong Wang}
\authornotemark[2]
\email{wangkl@hust.edu.cn}
\affiliation{%
  \institution{Huazhong University of Science and Technology}
  \city{Wuhan}           
  \country{China}
}

\author[P Gao]{Peiming Gao}
\email{peiming.gpm@mybank.cn}
\affiliation{%
  \institution{MYbank, Ant Group}
  \city{Hangzhou}           
  \country{China}
}

\author[Y Zhang]{Yuanchao Zhang}
\email{yuanchao.zhang@mybank.cn}
\affiliation{%
  \institution{MYbank, Ant Group}
  \city{Hangzhou}           
  \country{China}
}

\author[C Wei]{Chen Wei}
\authornotemark[3]
\email{juyi.wc@mybank.cn}
\affiliation{%
  \institution{MYbank, Ant Group}
  \city{Hangzhou}           
  \country{China}
}

\author[H Wang]{Haoyu Wang}
\authornotemark[2]
\email{haoyuwang@hust.edu.cn}
\affiliation{%
  \institution{Huazhong University of Science and Technology}
  \city{Wuhan}           
  \country{China}
}

\renewcommand{\shortauthors}{Jian Zhao et al.}

\begin{abstract}
The proliferation of pre-trained models~(PTMs) and datasets has led to the emergence of centralized model hubs like Hugging Face, which facilitate collaborative development and reuse. However, recent security reports have uncovered vulnerabilities and instances of malicious attacks within these platforms, highlighting growing security concerns. This paper presents the first systematic study of malicious code poisoning attacks on pre-trained model hubs, focusing on the Hugging Face platform. We conduct a comprehensive threat analysis, develop a taxonomy of model formats, and perform root cause analysis of vulnerable formats. While existing tools like \textsc{Fickling} and \textsc{ModelScan} offer some protection, they face limitations in semantic-level analysis and comprehensive threat detection. To address these challenges, we propose \textsc{MalHug}, an end-to-end pipeline tailored for Hugging Face that combines dataset loading script extraction, model deserialization, in-depth taint analysis, and heuristic pattern matching to detect and classify malicious code poisoning attacks in datasets and models. In collaboration with Ant Group, a leading financial technology company, we have implemented and deployed \textsc{MalHug} on a mirrored Hugging Face instance within their infrastructure, where it has been operational for over three months. During this period, \textsc{MalHug} has monitored more than 705K models and 176K datasets, uncovering 91 malicious models and 9 malicious dataset loading scripts. These findings reveal a range of security threats, including reverse shell, browser credential theft, and system reconnaissance. This work not only bridges a critical gap in understanding the security of the PTM supply chain but also provides a practical, industry-tested solution for enhancing the security of pre-trained model hubs.
\end{abstract}




\maketitle

\section{Introduction}
In recent years, Large Language Models~(LLMs) such as ChatGPT~\cite{gpt} have made significant progress, largely due to advancements in pre-training techniques. These pre-training methods have enabled the development of models with massive scale, often reaching billions or even trillions of parameters~\cite{lian2022trillion,nvidia2024trillion,william2022moe}. The reuse of these Pre-trained Models~(PTMs) has become increasingly important in advancing various AI applications. In this context, model hubs~(also known as model registries) like Hugging Face~\cite{huggingface} play a significant role in facilitating the reuse of pre-trained models~\cite{jiang2023huggingface}. Serving as a centralized repository, Hugging Face currently hosts an impressive collection of over 761K models and 176K datasets as of July 12, 2024~\cite{huggingface,jiang2024peatmoss}, which provides a collaborative environment for storing and sharing a wide variety of PTMs and datasets.

\noindent \textbf{Motivation.} With the emerging popularity and influence of model hubs, their centralized nature and widespread use also make them high-value targets for malicious actors~\cite{OWASP-LLM,jiang2022security,wang2024large}. Recent security reports have uncovered vulnerabilities~\cite{hiddenlayer2024hijacking,tencent2024datasets,legit2023aijacking,lasso2023apitoken} and instances of malicious attacks~\cite{jfrog2024malmodel,hiddenlayer2022pickle,hiddenlayer2023keras,hiddenlayer2023ransomware,trailofbits2024pickle2,blackhat2024pickle} within the Hugging Face platform, highlighting the growing security concerns in model hubs. One primary attack vector involves injecting malicious code into models~\cite{jfrog2024malmodel,hiddenlayer2023keras} or datasets~\cite{tencent2024datasets}. This can be achieved through various means, such as compromising developer accounts~\cite{lasso2023apitoken}, exploiting vulnerabilities in the platform's upload or verification processes~\cite{legit2023aijacking}, or disguising malicious code as legitimate model components~\cite{jfrog2024malmodel,trailofbits2024pickle1}. 
Of particular concern is the exploitation of certain serialization methods, such as Python's \texttt{pickle} module~\cite{pickledocument}, which have inherent security implications. This enables malicious actors to inject harmful code during the serialization process, which can then be executed when the compromised models are loaded for training or inference~\cite{trailofbits2021pickle}.
Malicious code poisoning can be used to achieve a range of nefarious goals, including but not limited to backdoor installation~\cite{jfrog2024malmodel,li2023malwukong,duan2021measuring}, sensitive information theft~\cite{trailofbits2024pickle1,guo2023empirical}, and ransomware deployment~\cite{hiddenlayer2023ransomware}.

\noindent \textbf{Research Gaps.} Security researchers are aware of these attacks and have proposed several defensive solutions. Trail of Bits has developed \textsc{Fickling}~\cite{trailofbits2021fickling}, a practical decompiler, static analyzer, and bytecode rewriter for \texttt{pickle} files. ProtectAI has introduced \textsc{ModelScan}~\cite{protectai2023modelscan}, a versatile tool designed to detect security issues across various model formats. Hugging Face has implemented \textsc{PickleScanning}~\cite{huggingface2024picklescanning}, which incorporates an anti-virus scan utilizing ClamAV~\cite{cisco2024clamav} and a targeted analysis that extracts and examines the list of imports referenced within \texttt{pickle} files.
While these solutions represent significant progress, they face notable limitations. These tools primarily rely on detecting specific libraries and function calls, rather than analyzing the actual executed code, which makes it challenging to conduct semantic-level analysis of malicious behaviors, potentially leading to both false positives and false negatives, especially when faced with sophisticated or obfuscated attacks. Moreover, there is a lack of comprehensive understanding of the abuse and attack techniques targeting the PTM supply chain. This gap in knowledge limits our ability to develop advancing defense strategies against the full spectrum of threats in PTM ecosystems.

% 在我们的研究中，首先进行了Empirical Study，对Hugging Face的风险面进行威胁建模，然后对目前主流的模型格式进行了大规模测量和分析，通过系统性的模型格式分类法来确定危险的模型文件格式并进行root cause分析。在此基础上，我们提出并设计了MalHug，这是一个为Hugging Face量身定制的自定义管道，通过结合元数据分析，模型反编译和静态污点分析来检测和分类Hugging Face上托管的恶意数据集与模型。

\noindent \textbf{Our Work.} Motivated by the above security concerns, we conduct the first systematic study of malicious code poisoning attacks on pre-trained model hubs, bridging the critical gap in understanding the vulnerabilities and attack vectors within the PTM supply chain. In our study, we first undertake a comprehensive pilot study, encompassing threat modeling, systematic model format taxonomy, and root cause analysis of vulnerable model formats. Building on these insights, we propose \toolname{}, an end-to-end pipeline tailored for Hugging Face that combines dataset loading scripts extraction, model deserialization, in-depth taint analysis, and heuristic pattern matching, enabling nuanced detection and classification of malicious datasets and models. 

% 我们实现了MalHug的原型系统，并持续监控Hugging Face上的超过174,000个数据集和751,000个模型，识别出XX个恶意数据集和XX个恶意模型，其中包括远程控制，敏感信息泄露等。

\noindent \textbf{Industrial Deployment.} We have implemented and deployed \toolname{} in collaboration with Ant Group, a leading financial technology company, demonstrating its scalability and effectiveness in a real-world industrial setting. \toolname{} has been operational for over three months on a mirrored Hugging Face instance within Ant Group's infrastructure, continuously monitoring more than 705K models and 176K datasets. Through this comprehensive industrial-grade analysis, \toolname{} has successfully identified 91 malicious models and 9 malicious dataset loading scripts, uncovering a range of security threats, including sophisticated remote control, browser credential theft, and system reconnaissance. These findings underscore the urgent need for robust security measures in industrial AI pipelines and provide valuable insights into the specific security challenges faced by large-scale financial technology companies in managing and deploying pre-trained models.

% We have implemented the prototype of \toolname{} and deployed it to continuously monitor over 751K models and 174K datasets on the Hugging Face platform. This extensive real-world evaluation demonstrates the scalability and effectiveness of our approach in a production environment. Through our comprehensive analysis, MalHug successfully identified X malicious datasets and Y malicious models, uncovering a range of security threats previously undetected by existing safeguards.
% The identified malicious content encompasses various types of threats, including remote control capabilities and sensitive information leakage. Specifically, we discovered Z instances of models containing code that could potentially establish unauthorized remote connections, posing significant risks to systems deploying these models. Additionally, W cases of datasets and models were found to leak sensitive information, such as API keys, credentials, or personal data, highlighting the critical need for robust security measures in the AI model ecosystem.

To summarize, we make the following contributions:

\begin{itemize}[leftmargin=15pt]
    \item \textbf{Systematic Study.} We conduct the first systematic study of malicious code poisoning attacks on PTM hubs, including comprehensive threat modeling, a systematic taxonomy of model formats, and root cause analysis of vulnerable model formats, which bridges a critical gap in understanding the vulnerabilities and attack vectors within the PTM supply chain.
    \item \textbf{Practical Pipeline.} We design and implement \toolname{}, an end-to-end pipeline tailored for Hugging Face. By integrating dataset loading script extraction, model deserialization, in-depth taint analysis, and heuristic pattern matching, \toolname{} offers a more nuanced and effective approach to detecting and classifying malicious PTMs and dataset loading scripts.
    \item \textbf{Real-world Impact.} \toolname{} has been operational for over three months on a mirrored Hugging Face instance within Ant Group's infrastructure, monitoring more than 705K models and 176K datasets. This analysis uncovered 91 malicious models and 9 malicious dataset loading scripts, providing valuable insights into securing the pipeline for managing and deploying PTMs. All these detected malicious artifacts have been made publicly available at \url{https://github.com/security-pride/MalHug}.
\end{itemize}
\section{Background}
In this section, we introduce the background of model hubs, present the threat model, and provide a taxonomy of model formats.

\subsection{Model Hubs and Artifact Reuse}
Model hubs, also known as model registries, have become integral to the AI ecosystem, serving as centralized repositories for pre-trained models, datasets, and associated resources. These platforms facilitate the distribution, discovery, and deployment of pre-trained models across various domains. \autoref{tab:model-hubs} presents an overview of the top 15 popular model hubs, showcasing the scale and diversity of available resources. Among these registries, Hugging Face \cite{huggingface} stands out as the largest and most comprehensive platform, hosting an impressive 752,269 models and 174,226 datasets as of July 6, 2024. 
% Its extensive collection spans a wide range of AI tasks and domains, making it a go-to resource for researchers, developers, and practitioners alike. 
Given its dominant position in the field and its significant impact, we have chosen to focus primarily on Hugging Face as the main subject of our study in this paper. 

\begin{table}[t]
    \centering
    \fontsize{8}{11}\selectfont
    \caption{Top 15 popular model hubs: number of models and datasets, and distribution mechanisms~(as of July 6, 2024). Note that ``-'' indicates no public statistics available or no dataset hosting service provided.}
    \begin{tabular}{crrc}
    \toprule
    \textbf{Model Hub} & \textbf{\#Models} & \textbf{\#Datasets} & \textbf{Distribution} \\
    \midrule
    Hugging Face~\cite{huggingfacemodels} & 752,269 & 174,226 & Hub APIs, Git \\
    Spark NLP~\cite{sparknlp} & 41,346 & -     & Hub APIs, Download \\
    OpenCSG~\cite{opencsg} & 26,187 & 327   & Git \\
    Kaggle~\cite{kaggle} & 5,932 & 355,251 & Hub APIs, Download \\
    ModelScope~\cite{modelscope} & 5,749 & 2,302 & Hub APIs, Git \\
    ModelZoo~\cite{modelzoo} & 3,245 & -     & Git \\
    OpenMMLab~\cite{openmmlab} & 2,404 & -     & Git \\
    ONNX Model Zoo~\cite{onnx_modelzoo} & 1,720 & -     & Git \\
    NVIDIA NGC~\cite{nvidia_ngc} & 759   & -     & Cli, Download \\
    MindSpore~\cite{mindspore} & 706   & 390   & Git, Download \\
    WiseModel~\cite{wisemodel} & 624   & 524   & Git \\
    PaddlePaddle~\cite{paddlehub} & 272   & 10,000 & Git \\
    SwanHub~\cite{swanhub} & 269   & -     & Git \\
    Liandanxia~\cite{liandanxia} & 264   & 381   & Git \\
    PyTorch Hub~\cite{pytorch} & 52    & -     & Hub APIs, Git \\
    \bottomrule
    \end{tabular}%
    \label{tab:model-hubs}%
\end{table}%

The proliferation of artifacts~(datasets and models) on Hugging Face has significantly impacted the landscape of AI research and development, fostering a culture of reuse and collaboration. Researchers and developers can leverage existing artifacts to train new models or fine-tune pre-trained ones for specific tasks, reducing the time and resources required for data collection, annotation, and model development. Hugging Face provides convenient tools for artifact reuse, such as libraries for loading datasets or accessing pre-trained models. For instance, users can easily load datasets using \texttt{datasets.load\_dataset()} function, and access pre-trained models via \texttt{AutoModel.from\_pretrained()} method. 
% This ecosystem of shared resources has led to the emergence of transfer learning as a dominant paradigm, where models pre-trained on large datasets are adapted for various downstream tasks using techniques like fine-tuning or prompt engineering.

\subsection{Code Poisoning Attacks on Model Hubs}

\noindent \textbf{Attack Vectors.} While model hubs like Hugging Face have greatly benefited the AI community, their centralized nature and widespread use also make them attractive targets for malicious actors~\cite{OWASP-LLM,jiang2022security,wang2024large}. To understand the security implications, we conduct a threat modeling and attack surface analysis of Hugging Face, focusing primarily on code poisoning attacks, which share similarities with supply chain attacks in open-source software ecosystems~\cite{ohm2020backstabber,duan2021measuring}. Recently, security researchers~\cite{tencent2024datasets,hiddenlayer2022pickle,jfrog2024malmodel} have reported two main attack vectors for code poisoning in model hubs:

\begin{itemize}[leftmargin=15pt]
    \item \textbf{Dataset Loading Scripts Exploitation.} Dataset loading script is a default feature provided by Hugging Face, typically employed to load datasets composed of data files in unsupported formats or requiring more complex data preparation. When users invoke the \texttt{load\_dataset} function, the corresponding loading script with the same name will be executed by default~\cite{huggingface2024dataset,huggingface2024loadingscripts}. While enhancing flexibility, this feature creates a significant attack surface, where malicious actors could embed harmful scripts within these datasets~\cite{tencent2024datasets}. 
    % When loaded, these scripts execute automatically, potentially causing arbitrary code execution~(ACE). 
    % This vulnerability exploits users' trust in reputable data sources and creates a substantial attack surface through automated script execution.

    \item \textbf{Insecure Model Serialization.} Many PTMs use insecure serialization formats like \texttt{pickle}~\cite{pickledocument}, which allow arbitrary code execution during deserialization. This creates a significant risk of injecting malicious code into model files. When users load compromised models, the embedded malicious code executes, potentially leading to severe security breaches~\cite{jfrog2024malmodel,hiddenlayer2022pickle}. 
    % This vulnerability is exacerbated by the widespread sharing of PTMs and users' implicit trust in models from popular repositories. 
    % The complexity and opaqueness of PTMs make detecting such embedded malicious code particularly challenging, rendering this attack vector difficult to defend against.
\end{itemize}

\noindent \textbf{Threat Model.} 
These attack vectors exploit the complex trust relationships within model hubs. To systematically analyze code poisoning attacks, we have developed a comprehensive threat model, which is based on several key assumptions. Firstly, users generally trust content from well-known model hubs and popular contributors, often prioritizing convenience and efficiency over rigorous security checks when using shared resources. Additionally, security measures on model hubs may not always keep pace with rapidly evolving threats. In this landscape, potential attackers have access to the public-facing interfaces of model hubs and can create and upload malicious datasets and models to these platforms. More concerning is their array of methods to gain or reinforce this trust within the community. For instance, attackers might exploit leaked authentication tokens~\cite{lasso2023apitoken} to gain unauthorized access to reputable accounts, allowing them to operate under the guise of trusted entities. They could also employ AI Jacking~\cite{legit2023aijacking} techniques, registering abandoned models or dataset names previously associated with respected organizations, thereby exploiting residual trust. These sophisticated approaches enable attackers to establish or hijack trusted identities within the model hubs, significantly increasing the potential impact of their malicious activities.

\begin{figure*}[t]
    \centering
    \includegraphics[width=0.9\linewidth]{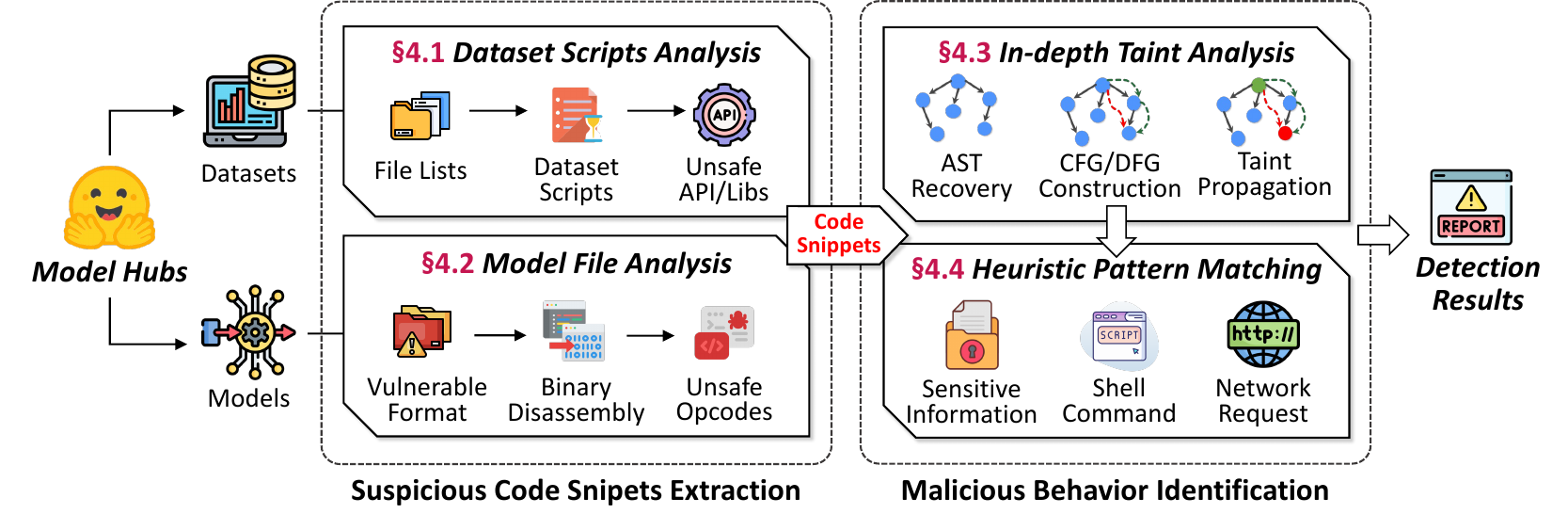}
    \caption{The workflow of \toolname{}: extracting suspicious codes from dataset loading scripts~(\autoref{sec:dataset}) and deserialized models~(\autoref{sec:model}), then applying taint analysis~(\autoref{sec:program}) and heuristic pattern matching~(\autoref{sec:rule}) to detect malicious behavior.}
    \label{fig:workflow}
\end{figure*}

\begin{table}[t]
\centering
\fontsize{8}{11}\selectfont
\caption{Taxonomy of 15 popular model formats and their vulnerability to code injection. 
Note that \CIRCLE ~indicates that this model format is vulnerable to code injection, \LEFTcircle ~represents partially vulnerable, and \Circle ~indicates that this model format is not vulnerable~(as of current knowledge).}
\begin{tabular}{>{\centering\arraybackslash}p{1.3cm}|ccc}
\hline
\textbf{Stored} & \textbf{Model Format} & \multicolumn{1}{c}{\textbf{Framework}} & \textbf{Injection?} \\
\hline
\multicolumn{1}{c|}{\multirow{12}[2]{*}{\parbox{1.3cm}{\centering Architecture \& Weights}}} 
  & pickle~\cite{pickledocument} & PyTorch, Scikit-learn & \CIRCLE \\
  & marshal~\cite{marshal2024} & /     & \CIRCLE \\
  & joblib~\cite{joblib2024} & PyTorch, Scikit-learn & \CIRCLE \\
  & dill~\cite{mckerns2012dill}  & PyTorch, Scikit-learn & \CIRCLE \\
  & cloudpickle~\cite{cloudpickle2024} & Scikit-learn, MLFlow & \CIRCLE \\
  & SavedModel~\cite{tensorflow2024savedmodel} & Tensorflow & \LEFTcircle \\
  & Checkpoint~\cite{tensorflow2024checkpoint} & TensorFlow & \LEFTcircle \\
  & TFLite~\cite{tensorflow2024tflite} & TFLite & \LEFTcircle \\
  & HDF5~\cite{hdfgroup2024hdf5}  & Keras & \LEFTcircle \\
  & GGUF~\cite{ggml2024gguf}  & llama & \Circle \\
  & ONNX~\cite{onnx2024}  & ONNX  & \Circle \\
\hline
\multicolumn{1}{c|}{\multirow{4}[2]{*}{\parbox{1.3cm}{\centering Weights Only}}} 
  & JSON~\cite{json2024}  & /     & \Circle \\
  & MsgPack~\cite{msgpack2024} & Flax  & \Circle \\
  & Safetensors~\cite{huggingface2024safetensors} & Huggingface & \Circle \\
  & NPY~\cite{npy2024}~/~NPZ~\cite{npz2024} & Numpy & \Circle \\
\hline
\end{tabular}%
\label{tab:taxonomy}%
\end{table}%

\section{Taxonomy and Root Cause Analysis}
Pre-trained models employ a diverse range of serialization formats for persistent storage and loading~\cite{protectai2023modelformats,trailofbits2024modelformats}. These formats can be categorized based on their serialization mechanisms, security implications, and prevalence in the PTM ecosystem. \autoref{tab:taxonomy} presents a comprehensive overview of 15 popular model formats, categorizing them based on their storage capabilities. The formats are broadly divided into two categories: those that store both architecture and weights, and those that store weights only.

\noindent \textbf{Formats Storing Both Architecture \& Weights.}
Formats that store both architecture and weights provide a complete representation of the model, including its structure and learned parameters. As shown in ~\autoref{tab:taxonomy}, several widely used formats in this category have varying levels of vulnerability to code injection attacks.

\begin{itemize}[leftmargin=15pt]
\item \textbf{Pickle Variants~(Insecure).} These Python-specific serialization formats, including pickle~\cite{pickledocument}, marshal~\cite{marshal2024}, joblib~\cite{joblib2024}, dill~\cite{mckerns2012dill}, and cloudpickle~\cite{cloudpickle2024}, are notorious for their susceptibility to code injection. They can execute arbitrary Python code during deserialization, making them highly vulnerable when handling untrusted data.

\item \textbf{TensorFlow and Keras Models~(Potential).} These formats, primarily associated with TensorFlow~\cite{tensorflow2024savedmodel,tensorflow2024checkpoint,tensorflow2024tflite} and Keras~\cite{hdfgroup2024hdf5}, have a reduced but still present attack surface. They support custom operators~(SavedModel, Checkpoint, TFLite)~\cite{tensorflow2024security} or Lambda layers~(HDF5)~\cite{hiddenlayer2023keras} that can potentially execute arbitrary code, though with some additional barriers compared to pickle-like formats.

\item \textbf{GGUF and ONNX~(Secure).} These more recent formats show promise in terms of security, with no known vulnerabilities to code injection as of current knowledge. They strictly limit their scope to predefined model computation and transformation operations, avoiding support for arbitrary code execution or object instantiation~\cite{protectai2023modelformats}.
\end{itemize}

\find{\textbf{ Root Cause }$\blacktriangleright$
The vulnerabilities in these formats stem from a fundamental tension between flexibility and security in serialization design. 
Arbitrary object instantiation in pickle variants creates the most severe security risk, effectively blurring the line between data and code.
Lambda layers, particularly in HDF5~(Keras), introduce an indirect but significant risk through their dependency on the marshal module.
Custom operators in formats like SavedModel and TFLite present a smaller attack surface, as they require explicit loading during inference, but still pose potential risks.
$\blacktriangleleft$ }

\noindent \textbf{Formats Storing Weights Only.}
Formats that store only weights provide a more focused representation of the model, containing just the learned parameters without the architectural details. 
% This approach can offer improved security against code injection attacks, as the architecture is typically defined separately in code. 
As shown in \autoref{tab:taxonomy}, these formats generally have a lower risk of code injection vulnerabilities.

\begin{itemize}[leftmargin=15pt]
\item \textbf{JSON~(Secure).} While not specifically designed for PTM storage, JSON~\cite{json2024} can be used to store model weights. JSON is generally safe from code injection as it only supports basic data types and structures, without the ability to represent code or complex objects.

\item \textbf{MsgPack~(Secure).} MessagePack~(MsgPack)~\cite{msgpack2024} is a binary serialization format. 
% It's similar to JSON in terms of the data it can represent but more compact. 
MsgPack doesn't support code serialization, making it resilient against direct code injection attacks.

\item \textbf{Safetensors~(Secure).} Developed by Hugging Face~\cite{huggingface2024safetensors}, Safetensors could prevent code injection attacks. It uses a simple, language-agnostic format that strictly limits deserialization to numerical data, effectively eliminating the risk of arbitrary code execution during the loading process.

\item \textbf{NPY / NPZ~(Secure).} These NumPy-specific formats~\cite{npy2024,npz2024} are primarily designed for storing numerical arrays. While they don't directly support code execution during deserialization, care must be taken to properly handle the data to avoid potential buffer overflow vulnerabilities.
\end{itemize}

\find{\textbf{ Security Features }$\blacktriangleright$
The security advantages of these formats highlight the importance of separating model architecture (which may require more complex serialization) from weight storage, especially when dealing with potentially untrusted data sources.
$\blacktriangleleft$ }

\section{\toolname{} Workflow}

% 以上的威胁建模和分类研究表明，在Hugging Face上进行Code Poisoning Attack是相当容易的。尽管以往的安全研究部分报告了in the wild的模型投毒，并且有一些检测工具被开发出来，但仍然缺乏一个系统性的研究和端到端的分析框架。在本章中，我们介绍了\toolname{}，一个针对Hugging Face的端到端的pipeline。我们在图1中展示了\toolname{}的工作流程，它由四个组件组成：数据集预处理，模型预处理，深层污点分析，和启发式模式匹配，最终揭示注入到数据集脚本或者脆弱模型文件中的恶意代码。

In this section, we introduce \toolname{}, a comprehensive end-to-end pipeline specifically designed for Hugging Face, focusing on detecting code poisoning attacks on dataset loading scripts and vulnerable models files~(Pickle variants and lambda layers in HDF5). \autoref{fig:workflow} illustrates the workflow of \toolname{}, which comprises four key components: dataset loading scripts extraction, model deserialization, in-depth taint analysis, and heuristic pattern matching. 
% Notably, \toolname{} does not aim to invent new program analysis techniques. 
% \toolname{} leverages insights from existing attacks to construct an efficient and practical review pipeline for identifying and analyzing potential security threats within Hugging Face.

\subsection{Dataset Loading Scripts Analysis}
\label{sec:dataset}
The dataset pre-processing forms the initial step of our pipeline, focusing on the extraction and examination of loading scripts associated with datasets from Hugging Face.

\begin{table}[t]
\centering
\fontsize{8.5}{12}\selectfont
\caption{Unsafe libraries and APIs.}
\label{tab:risky-apis-libs}
\begin{tabular}{cc}
\hline
\textbf{Category} & \textbf{Unsafe Libs/APIs} \\
\hline
\multirow{3}{*}{Builtin Functions} & \texttt{eval, exec, execfile} \\
& \texttt{\_\_import\_\_, getattr} \\
& \texttt{compile, open} \\
\hline
\multirow{2}{*}{Command Execution} & \texttt{os.system/popen/spawn*} \\
& \texttt{subprocess.run/call/Popen} \\
\hline
\multirow{4}{*}{Network} & \texttt{requests.get/post} \\
& \texttt{urllib.request.urlopen/Request} \\
& \texttt{socket.socket/connect} \\
& \texttt{ftplib.FTP, smtplib.SMTP} \\
\hline
\multirow{4}{*}{File System} & \texttt{shutil.rmtree/move} \\
& \texttt{pathlib.Path, os.path.join} \\
& \texttt{zipfile.ZipFile, tarfile.open} \\
& \texttt{glob.glob, fnmatch.filter} \\
\hline
\multirow{2}{*}{System Information} & \texttt{os.environ/getcwd} \\
& \texttt{platform.system/release} \\
\hline
\multirow{4}{*}{Cryptography} & \texttt{Crypto.Cipher.AES/DES} \\
& \texttt{cryptography.fernet.Fernet} \\
& \texttt{rsa.encrypt/decrypt} \\
& \texttt{base64.b64encode/b64decode} \\
\hline
\end{tabular}
\end{table}

\noindent \textbf{Unsafe Library and API Filtering.} We begin by extracting the loading script associated with each dataset obtained from Hugging Face. Once the relevant scripts are extracted, we perform an initial analysis to identify unsafe libraries and APIs. This process involves scanning the script contents for import statements and function calls and cross-referencing them against a curated list of potentially unsafe libraries and APIs. To ensure a comprehensive and accurate review, we synthesize the static analysis rules used in Pyre~\cite{pyre} and Semgrep~\cite{semgrep}, thereby compiling a more extensive list of insecure libraries and APIs, as shown in \autoref{tab:risky-apis-libs}. The risky Libraries and APIs including known dangerous functions (e.g., \texttt{eval}, \texttt{exec}), libraries associated with command execution (e.g., \texttt{os}, \texttt{subprocess}), and networking modules that could indicate unauthorized data transmission (e.g., \texttt{requests}, \texttt{urllib}). We employ regular expressions and AST~(Abstract Syntax Tree) parsing to efficiently identify these elements within the code. 

% 根据Section3中对脆弱模型格式的分析，我们主要关注于Pickle Variants, Keras Models和TensorFlow Models。首先我们确认脆弱模型的格式，然后根据不同的模型格式执行二进制反汇编。
% \noident \textbf{Pickle Scan.} 对于.pth、.pt或.bin格式保存的PyTorch模型文件，它们实际上是一个ZIP压缩包，其中通常包含一个data.pkl的权重文件。反汇编通过pickletools进行，将二进制反汇编为Opcode，并检查其中Unsafe的Opcode。如果存在unsafe的opcode，\toolname{}会进一步利用Fickling反编译获取AST，然后提取其中注入的可疑代码片段，以供后续in-depth taint analysis。如示例所示。
% \noident \textbf{Keras Scan.} 对于keras或h5保存的Keras模型文件，\toolname{}检查其中的lambda层调用
% \noident \textbf{TensorFlow Scan.} 对于pb, ckpt, tflite保存的TensorFlow模型，\toolname{}检测其中的文件读写操作符。

\subsection{Model File Analysis}
\label{sec:model}

Model deserialization is a crucial step in our security analysis pipeline, designed to uncover potentially malicious code or suspicious operations within model files. Our approach is tailored to handle various vulnerable model formats used by popular frameworks such as PyTorch, Keras, and TensorFlow.

\begin{figure*}[t]
    \centering
    \includegraphics[width=0.85\linewidth]{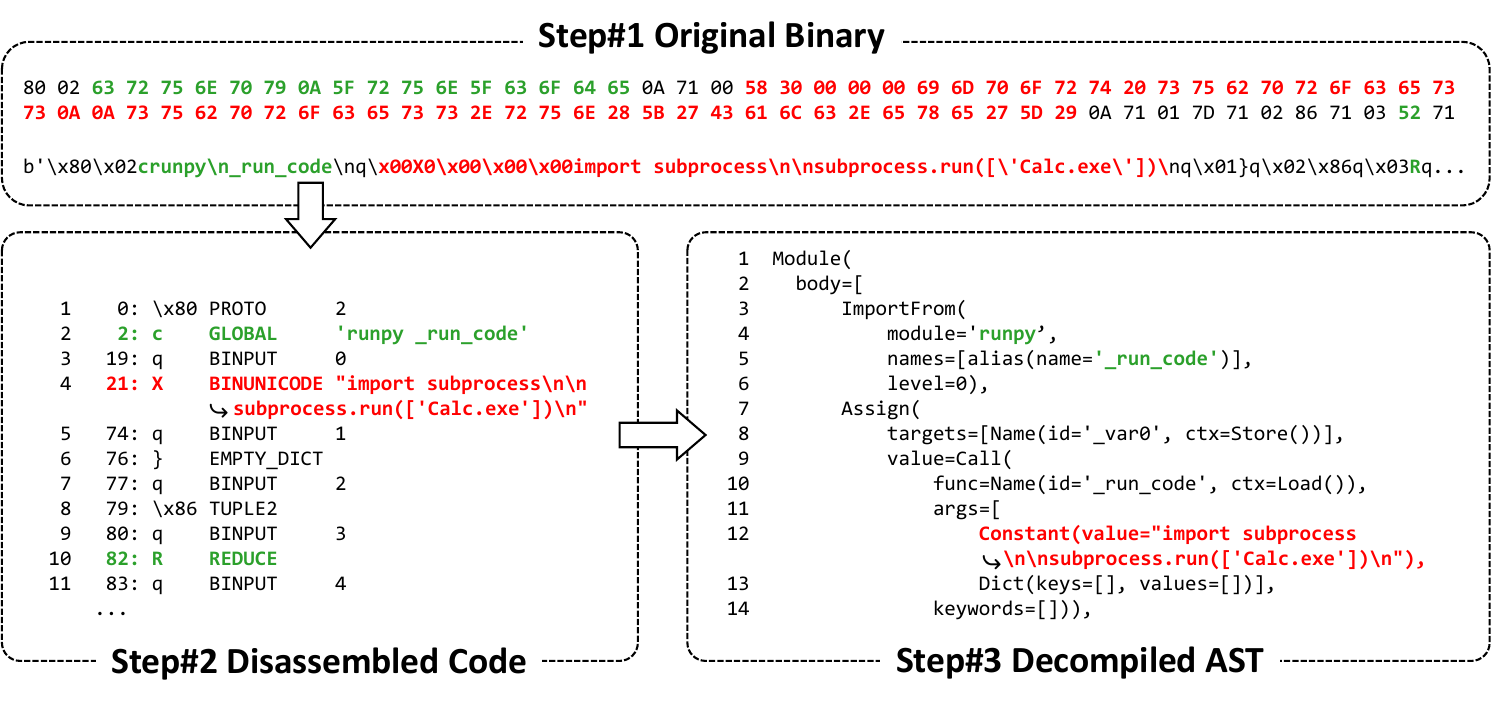}
    \caption{The Pickle model decompilation process of \texttt{MustEr/gpt2-elite}. Snippet~\#1 is the original binary code, snippet~\#2 is the disassembled code, and snippet~\#3 is the decompiled AST. \textcolor{codegreen}{Green} highlights suspicious opcodes, while \textcolor{codered}{Red} indicates potentially malicious injected code.}
    \label{fig:code-snippet}
    % \vspace{2em}
\end{figure*}

\noindent \textbf{PyTorch/Pickle Variants.} 
For PyTorch models saved in \texttt{.pth}, \texttt{.pt}, or \texttt{.bin} formats, which are essentially ZIP archives typically containing a \texttt{data.pkl} weights file, we employ a multi-stage decompilation process to analyze potentially malicious code without execution risk. As illustrated in \autoref{fig:code-snippet}, our process begins with extracting the \texttt{data.pkl} file from the model archive~(Step\#1). We then use \texttt{pickletools}\cite{python2024pickletools} to disassemble the pickle bytecode into human-readable opcodes~(Step\#2). This disassembly reveals the underlying structure of the serialized data, such as the \texttt{GLOBAL} opcode~(Step\#2, line 2), which imports the \texttt{runpy.\_run\_code} function, a potential vector for code execution. Through a systematic manual audit of all opcodes mentioned in \texttt{pickle}~\cite{picklefile}, we identify and summarize the potentially unsafe opcodes associated with code execution. The results of this analysis are presented in \autoref{tab:unsafe-pickle-opcodes}. We scan these opcodes for unsafe operations that could lead to code injection.
Upon detecting such unsafe opcodes, we employ \textsc{Fickling}\cite{trailofbits2021fickling} to further decompile the pickle file into an AST, as depicted in Step\#3. This higher-level representation exposes the structure of the potentially malicious code. From the AST, we extract suspicious code snippets by analyzing function call arguments. In \autoref{fig:code-snippet}, we identify a function call to \texttt{runpy.\_run\_code} with a constant argument that appears to be a Python script~(Step\#3, line 10-12), which is extracted as potentially malicious code.

\begin{table}[t]
\centering
\fontsize{9}{12}\selectfont
\caption{Unsafe pickle opcodes.}
\label{tab:unsafe-pickle-opcodes}
\begin{tabular}{cl}
\hline
\textbf{Opcode} & \textbf{Description} \\
\hline
\texttt{REDUCE} & Applies callable object to argument tuple \\
\texttt{(b`R')} & Pops function and args, pushes return value \\
\hline
\texttt{GLOBAL} & Imports modules or gets global objects \\
\texttt{(b`c')} & Pushes retrieved object onto stack \\
\hline
\texttt{OBJ} & Builds class instance (Protocol 1) \\
\texttt{(b`o')} & Uses class object from stack \\
\hline
\texttt{INST} & Builds class instance (Protocol 0) \\
\texttt{(b`i')} & Uses module and class names \\
\hline
\texttt{NEWOBJ} & Builds object instance using \texttt{\_\_new\_\_} \\
\texttt{(b`\textbackslash x81')} & Calls cls.\_\_new\_\_(cls, *args) \\
\hline
\texttt{NEWOBJ\_EX} & Extended version of NEWOBJ \\
\texttt{(b`\textbackslash x92')} & Calls cls.\_\_new\_\_(cls, *args, **kwargs) \\
\hline
\end{tabular}
\end{table}

\noindent \textbf{TensorFlow/Keras Model.} 
The process of deconstructing and analyzing TensorFlow and Keras models, as outlined in \autoref{alg:tensorflow-and-keras}, focuses on detecting Lambda layers and unsafe operators within these models. This process begins with \texttt{ParseModelStructure}~(line 1), which handles two primary formats: SavedModel and HDF5. For SavedModel, we utilize \texttt{SavedMetadata.ParseFromString}~\cite{blackhat2024confused} to load the model metadata and \texttt{SavedModel.ParseFromString}~\cite{so2020savedmodel} to load the model itself. For HDF5 format, we employ \texttt{h5py.File}~\cite{blackhat2024confused,h5py2024file} to read the model file,  extracting \texttt{model\_config} attribute containing a JSON string of the model architecture, and parsing this JSON string to obtain layer configurations.
Once the model structure is parsed, our algorithm iterates through each layer using \texttt{IterateLayers}~(lines 7-15). This function abstracts the differences between SavedModel and HDF5 formats, providing a unified interface for layer iteration. During iteration, we check for \texttt{Lambda} layers using \texttt{IsLambdaLayer}. Simultaneously, we employ another funcion \texttt{CheckForUnsafeOperators}~(lines 16-22) to identify any usage of potentially risky operations. This function searches for specific TensorFlow operations that could pose security risks, such as file I/O operations~(\texttt{tf.io.read\_file}~\cite{tensorflow2024readfile}, \texttt{tf.io.write\_file}~\cite{tensorflow2024writefile}).
For Keras models with Lambda layers, we decompile the Python bytecode stored in the marshal-serialized format. By adding appropriate Python version headers to the Lambda layer data, we can leverage a rich set of \texttt{.pyc} decompilation tools~\cite{pyc2src,pyc2src2,pyc2src3} to obtain equivalent Python source code snippets, which allows us to examine the content of Lambda layers more thoroughly. However, for unsafe operators in TensorFlow models, we do not perform further analysis beyond identification. This decision is based on the fact that these operators cannot directly inject system commands. As a result, TensorFlow models using unsafe operators are simply flagged as potentially unsafe without undergoing additional examination.

% \shenao{Update this!} Note that we do not conduct in-depth taint analysis on Lambda layers or unsafe operators. This decision is based on several technical constraints and considerations. Primarily, the deserialization of marshal-encoded Lambda layers depends on \texttt{marshal.load}~\cite{marshal2024} of specific Python versions. Moreover, these layers would require dynamic loading in a sandboxed environment. Given these limitations, we have opted to simply flag these instances as potentially unsafe and manually check these models.

\begin{algorithm}[t]
\fontsize{8.5}{9}\selectfont
\caption{Unsafe Keras/TensorFlow Model Detection.}
\label{alg:tensorflow-and-keras}
\SetKwInput{KwInput}{Input}
\SetKwInput{KwOutput}{Output}
\SetKwProg{Fn}{Function}{:}{}

\KwInput{Model file $M$, a set of $unsafe\_opt$,}
\KwOutput{Usage of Lambda layers and unsafe operators}

$model \gets \text{ParseModelStructure}(M)$\;
\ForEach{$layer \in \text{IterateLayers}(model)$}{
    \If{$\text{IsLambdaLayer}(layer)$}{
        $has\_lambda\_layer \gets \text{True}$\;
        \textbf{break}\;
    }
    $unsafe\_opt.\text{update}(\text{CheckForUnsafeOpt}(layer))$\;
}
\Fn{IterateLayers($model$)}{
    \If{$model$ is SavedMetadata}{
        \ForEach{$node \in model.nodes$}{
            \If{$node.identifier = \text{``\_tf\_keras\_layer''}$}{
                $layer \gets \text{JSON.parse}(node.metadata)$\;
                \textbf{yield} $layer$\;
            }
        }
    }{
        $config \gets \text{parse}(model.attrs[\text{``model\_config''}])$\;
        \ForEach{$layer \in config[\text{``config''}][\text{``layers''}]$}{
            \textbf{yield} $layer$\;
        }
    }
}
\Fn{CheckForUnsafeOpt($layer$)}{
    $unsafe\_ops \gets \text{Set}()$\;
    $risky\_ops \gets [\text{``tf.io.read\_file''}, \text{``tf.io.write\_file''}]$\;
    \ForEach{$op \in risky\_ops$}{
        \If{$\text{contains}(layer.\text{to\_string}(), op)$}{
            $unsafe\_ops.\text{add}(op)$\;
        }
    }
    \Return $unsafe\_ops$\;
}
\Return $has\_lambda\_layer, unsafe\_opt$\;
\end{algorithm}

\subsection{In-depth Taint Analysis}
\label{sec:program}

After extracting suspicious code snippets from dataset loading scripts and model files~(PyTorch \& Keras), \toolname{} implements a focused taint analysis, which has been proven to be good at detecting a wide range of malicious code poisoning attack patterns in previous studies~\cite{li2023malwukong,duan2021measuring}. To perform this analysis, we build \toolname{} on an open-source static analysis framework \textsc{Scalpel}\cite{li2022scalpel}. We use \textsc{Scalpel} to construct control flow and data flow graphs, which serve as the foundation for our taint analysis. 
On top of this foundation, we define a comprehensive taint configuration based on a categorized set of source and sink APIs. These APIs are typically drawn from the unsafe APIs listed in \autoref{tab:risky-apis-libs}, but we assign them to specific source-sink combinations based on different malicious behavior patterns. Our configuration encompasses a wide range of potential security threats, including hidden authentication, backdoors, cryptojacking, embedded shells, remote control, sensitive information leakage, and suspicious execution patterns.

For each category of threat, we identify specific classes of source and sink APIs that could indicate malicious behavior. 
For example, in the case of sensitive information leakage attempts, we might consider \texttt{os.envirion} or \texttt{os.getlogin} as sources, and \texttt{requests.get} or \texttt{socket.connect} as sinks. This combination could reveal attempts to collect sensitive system information and transmit it to an unauthorized external server. 
For remote control attempts detection, we might consider the reverse shell commands as sources, and APIs from the command execution as sinks, such as \texttt{os.system}, \texttt{os.spawn*}, and \texttt{subprocess.run}, possibly indicating the injection of unauthorized shell commands. These source-sink pairings allow us to track the flow of potentially malicious operations through the code, providing a nuanced understanding of various attack vectors.

% In the context of cryptojacking, executing binary files could be a source, while \texttt{os.system()} or \texttt{subprocess.Popen()} might serve as sinks, possibly indicating the launch of unauthorized mining processes.

\subsection{Heuristic Pattern Matching}
\label{sec:rule}

While our taint analysis provides a robust framework for detecting malicious behaviors based on API and library usage, we recognize that not all sources of potential threats can be defined solely through Python APIs or libraries. Certain taint sources, such as malicious shell commands or obfuscated malicious code patterns, cannot be effectively marked through API-based methods alone. To address this limitation and enhance our detection capabilities, we incorporate heuristic pattern matching as a complementary technique to our taint analysis approach, leveraging YARA~\cite{vt2024yara} rules for efficient and flexible pattern matching.
% By applying these YARA-based heuristics, we can detect various sophisticated attack patterns that do not rely solely on API calls, including reverse shell commands potentially obfuscated within strings or comments. 
This dual-pronged strategy significantly enhances our ability to identify both API-based and pattern-based threats, enabling \toolname{} to achieve a more comprehensive and nuanced detection of malicious code in pre-trained models.

\section{Evaluation}

\subsection{Experimental Setup}
% MalHug在的模型反编译模块在开源Fickling和ModelScan的基础上构建，初步过滤可疑模型。此外，在scalpel的基础上实现in-depth污点分析，并基于自定义的YARA规则来检测恶意的污点数据流模式。
\noindent \textbf{Implementation.}
We have implemented a prototype of \toolname{} and deployed it on the mirrored Hugging Face instance within Ant Group for over three months. The model decompilation module of \toolname{} is built upon the open-source \textsc{Fickling}~\cite{trailofbits2021fickling} and \textsc{ModelScan}~\cite{protectai2023modelscan}, enabling preliminary filtering of suspicious models. Furthermore, \toolname{} implements in-depth taint analysis based on the \textsc{Scalpel}~\cite{li2022scalpel}, complemented by custom YARA~\cite{vt2024yara} rules to detect malicious taint flow patterns.

\noindent \textbf{Environment.}
The prototype of \toolname{} runs on a server with Ubuntu Linux 22.04, equipped with two AMD EPYC Milan 7713 CPUs (2.0 GHz, 64 cores, 128 threads each), 512 GB RAM (8 x 64 GB modules), two NVIDIA A100 GPUs with 80 GB memory each, and four 7.68 TB NVMe SSDs (Western Digital SN640), providing a total storage capacity of 30.72 TB. The Hugging Face mirror synchronization service runs on an Alibaba Cloud ECS instance (\texttt{ecs.c6a.16xlarge}), optimized for data-intensive storage operations. The server operates on Alibaba Cloud Linux 3 and is equipped with 64 vCPUs, 128 GB of RAM, and 8 data disks, each with 32 TB capacity, providing a total storage of 256 TB.

\noindent \textbf{Dataset.}
Due to the current lack of high-quality ground truth datasets of malicious artifact samples, we aim to evaluate the performance of \toolname{} in the real world and conduct a comprehensive investigation and measurement of code poisoning attacks in the real world. We download and detect accessible artifacts~(models and datasets) on the largest model hosting platform, Hugging Face. Specifically, we use Hugging Face's official Python library, \texttt{huggingface-hub}~\cite{huggingface2024hfapi}, to automatically collect metadata of 760,999 models and 176,849 datasets as of July 12. After excluding models with restricted access permissions, we conduct a comprehensive analysis of 705,991 models and 176,386 datasets, collectively amounting to 179.4 TB of data.
% Considering the potential resource consumption of downloading, storing, and analyzing large models, we filtered models smaller than 1 GB to ensure feasibility.

\subsection{Industrial Deployment \& Measurement}
\noindent \textbf{Vulnerable Dataset Loading Scripts.} Among the 176,386 mirrored datasets, 6,578~(3.73\%) contain loading scripts. These scripts play a crucial role in data preprocessing pipelines, potentially introducing security vulnerabilities and compromising the integrity of AI workflows if not properly scrutinized. Subsequently, \toolname{} focuses its main analysis on the code within these 6,578 dataset loading scripts to identify and assess potential security risks.

\begin{figure}[t]
    \centering
    \includegraphics[width=1\linewidth,keepaspectratio]{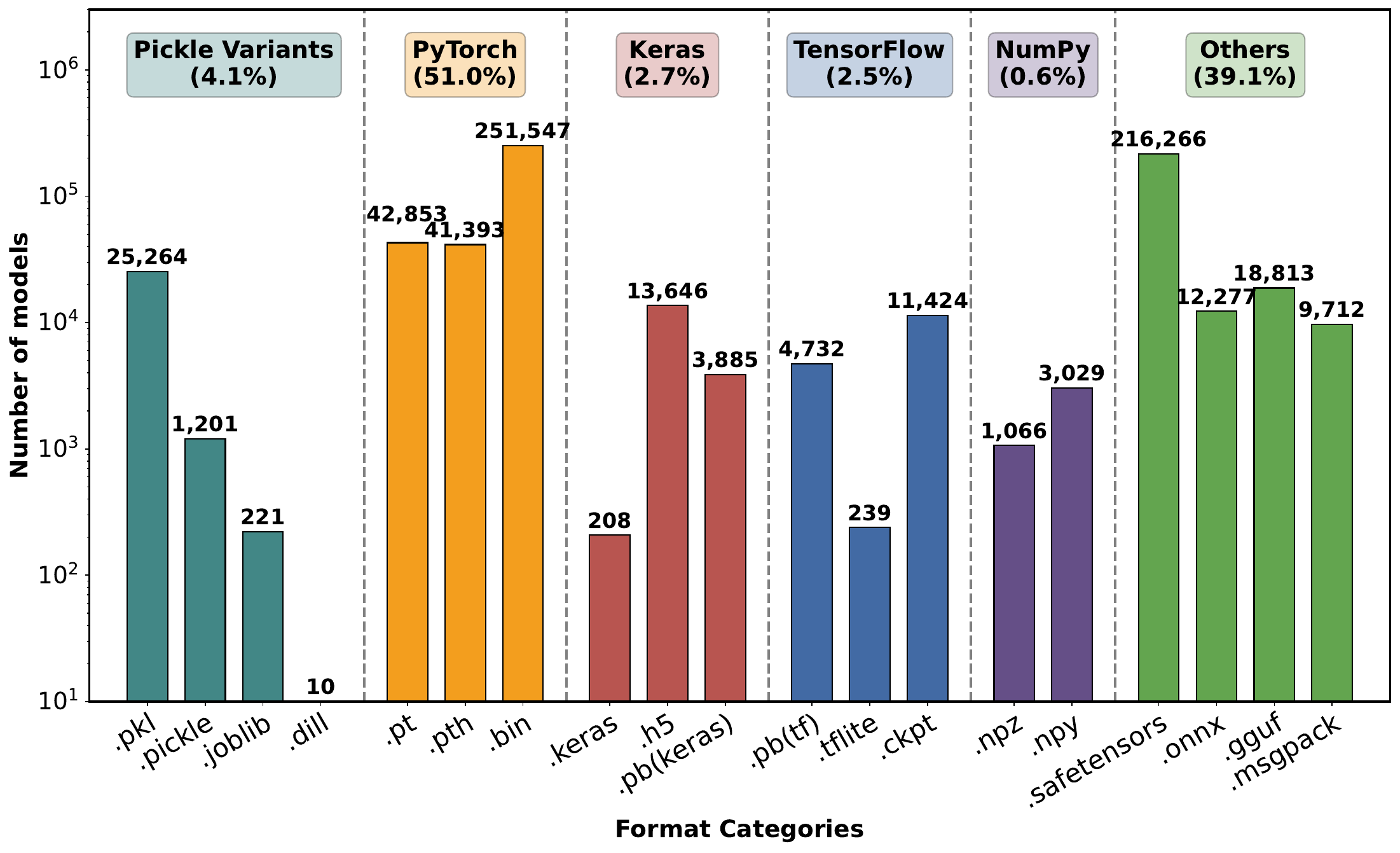}
    \caption{Distribution of model file formats in Hugging Face.}
    \label{fig:extension_usage}
\end{figure}

\noindent \textbf{Vulnerable Model Files.} Our investigation covers 705,991 mirrored model repositories, of which 133,058 are empty~(containing only \texttt{.gitattributes} and \texttt{README.md}). Among non-empty repositories, we observe a diverse range of model formats, as illustrated in \autoref{fig:extension_usage}, with a significant portion potentially vulnerable to security risks. PyTorch models~(\texttt{.pt/.pth/.bin}), which fundamentally use Pickle for serialization, are most prevalent with 335,893~(51.0\%) instances. This, combined with explicit Pickle variants (\texttt{.pkl, .pickle, .joblib, .dill}) accounting for 26,696~(4.1\%) models, means that over 55\% of the models use Pickle-based serialization, raising substantial security concerns.
Additional vulnerable formats include Keras models (\texttt{.keras/.h5/.pb}, with 17,739 (2.7\%) instances, and TensorFlow models (\texttt{.pb/.tflite/.ckpt}, accounting for 16,395 (2.49\%) of the total. This distribution highlights the critical need for comprehensive security measures across various serialization methods, particularly given the widespread use of potentially vulnerable formats like Pickle-based serialization in PyTorch models. Note that each model repository may contain multiple model formats, explaining why the total number of models exceeds the number of repositories.

% Other formats include \texttt{.safetensors} (216,266, 32.3\%) and \texttt{.onnx}~(12,277, 1.9\%), the latter designed as a more secure alternative.

\begin{figure*}[t]
    \centering
    \includegraphics[width=0.95\linewidth]{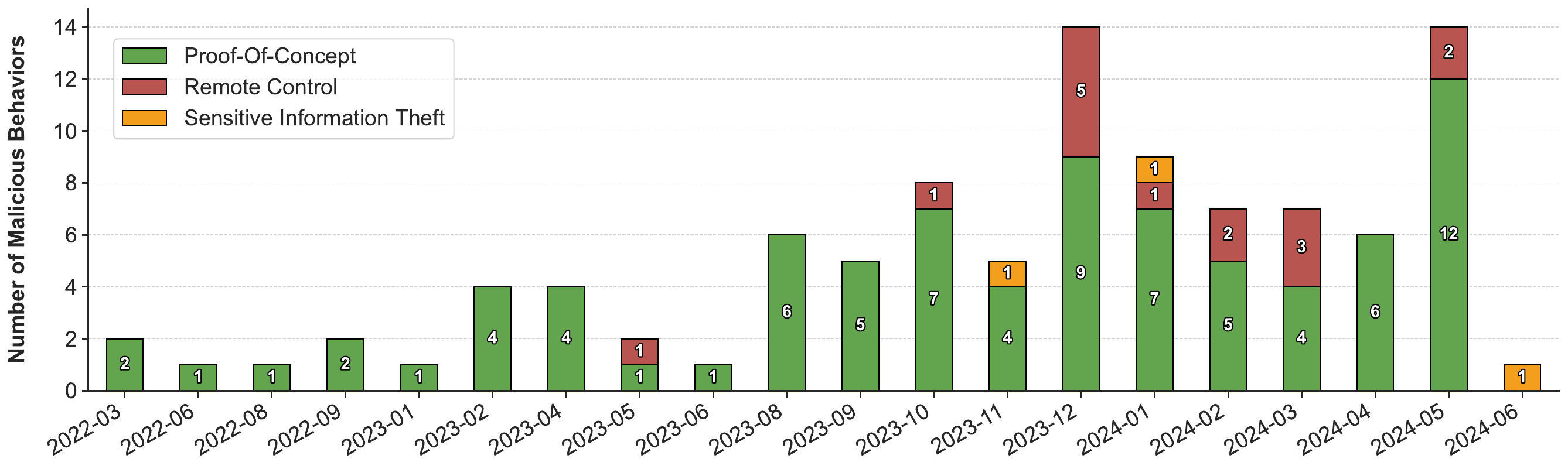}
    \caption{Monthly distribution and classification of malicious behaviors in models and dataset loading scirpts.}
    \label{fig:monthly_malicious_behaviors}
       \vspace{1em}
\end{figure*}

\begin{table}[t]
\centering
\fontsize{8.5}{11}\selectfont
\caption{Partial results of main unsafe Libs/APIs filtering.}
\label{tab:unsafe-api-res}
\begin{tabular}{c|cr}
\hline
\multicolumn{3}{c}{\textbf{Models}} \\
\hline
\textbf{Format/Type} & \textbf{API} & \textbf{\#Cnt} \\
\hline
\multirow{7}{*}{Pickle} & \_\_builtin\_\_.exec & 27 \\ 
& \_\_builtin\_\_.compile & 1 \\ 
& \_\_builtin\_\_.eval & 23 \\ 
& \_\_builtin\_\_.getattr & 3,775 \\ 
& runpy.\_run\_code & 6 \\ 
& os.system/posix.system & 18 \\ 
& webbrowser.open & 3 \\ 
\hline
Keras & Lambda & 72 \\ 
\hline
TensorFlow & ReadFile/WriteFile/etc. & 35 \\ 
\hline
\multicolumn{3}{c}{\textbf{Dataset Loading Scripts}} \\
\hline
YAML & yaml.load & 1 \\ 
\hline
\multirow{5}{*}{Eval and Execution} & \_\_builtin\_\_.compile & 56 \\ 
& \_\_builtin\_\_.eval & 74 \\ 
& \_\_builtin\_\_.getattr & 456 \\ 
& \_\_builtin\_\_.\_\_import\_\_ & 3 \\ 
& \_\_builtin\_\_.exec & 2 \\ 
\hline
\multirow{2}{*}{Command Execution} & os.system & 12 \\ 
& subprocess.* & 13 \\ 
\hline
\multirow{3}{*}{Network} & urllib.request.* & 32 \\ 
& urllib.parse.* & 15 \\ 
& aiohttp.client.get & 1 \\ 
\hline
\multirow{3}{*}{Cryptography} & base64.b64encode & 5 \\ 
& base64.urlsafe\_b64encode & 1 \\ 
& base64.b64decode & 8 \\ 
\hline
\textbf{Total} & \textbf{/} & \textbf{4,639} \\ %3960 + 679
\hline
\end{tabular}
\end{table}

\noindent \textbf{Unsafe API Filtering.} Our comprehensive analysis reveals the distribution of suspicious APIs across models and dataset loading scripts, as shown in \autoref{tab:unsafe-api-res}. 
In model files, we observe 27 occurrences of \texttt{\_\_builtin\_\_.exec}, 23 of \texttt{\_\_builtin\_\_.eval}, and 18 instances of \texttt{os.system} or \texttt{posix.system}. Dataset loading scripts exhibit a higher frequency of eval and execution functions, with 56 cases of \texttt{\_\_builtin\_\_.compile} and 74 of \texttt{\_\_builtin\_\_.eval}. Notably, the \texttt{getattr} function is overwhelmingly used despite Huggingface's clear ``unsafe'' label, accounting for 91.2\% of dangerous API usage (3,775 instances in models and 456 in dataset loading scripts). Upon closer inspection of the parameters passed to \texttt{getattr}, we do not identify any instances of actual malicious exploitation. While \texttt{getattr} can potentially be used to dynamically access sensitive or dangerous functions, its application in these contexts appears to be largely for legitimate programming purposes.
Additionally, we find 72 instances of Keras \texttt{Lambda} layers and 35 cases of unsafe TensorFlow operators, which will undergo further inspection to confirm their safety.
% Importantly, the identification of these unsafe APIs is not conclusive evidence of malicious intent, but rather a starting point for further investigation. To thoroughly assess the actual threat posed by these suspicious API calls, we conduct in-depth taint analysis on the flagged instances to identify malicious behaviors. 
% This detailed examination allowed us to analyze the malicious behavior patterns, if any, and distinguish between genuine security threats, necessary research-related usage~(such as in security proof-of-concept experiments), and potential false positives.

\begin{table*}[t]
\centering
\fontsize{9}{12}\selectfont
\caption{Qualitative comparison with other SOTA techniques.}
\begin{tabular}{ccccc}
\toprule
\textbf{Tools} & \textbf{Developer} & \textbf{Granularity} & \textbf{Dataset Support?} & \textbf{Model Format Support?} \\
\midrule
\textsc{Pickle Scanning}~\cite{huggingface2024picklescanning} & HuggingFace & Unsafe Lib~\&~API  & \xmark    & Pickle Only \\
\textsc{PickleScan}~\cite{mmaitre2024picklescan} & mmaitre314 & Unsafe Lib~\&~API  & \xmark    & Pickle Only \\
\textsc{Fickling}~\cite{trailofbits2021fickling} & Trail of Bits & Unsafe Lib~\&~API  & \xmark    & Pickle Only \\
\textsc{Bhakti}~\cite{dropbox2024bhakti} & Dropbox Inc & Unsafe Lib~\&~API  & \xmark    & Tensorflow~\&~Keras  \\
\textsc{ModelScan}~\cite{protectai2023modelscan} & ProtectAI & Unsafe Lib~\&~API  & \xmark    & Pickle Variants; Tensorflow~\&~Keras \\
\hline
\toolname{} & / & Semantic Level & \cmark   & Pickle Variants; Tensorflow~\&~Keras \\
\bottomrule
\end{tabular}%
\label{tab:qualitative}%
\end{table*}%

% \begin{figure}[t]
%   \centering
%   \subfloat[Quarterly Distribution.]{\includegraphics[width=1.8in]{figures/distribution_quarterly.pdf}\label{fig:quarterly}}
%   \subfloat[Types Distribution.]{\includegraphics[width=1.8in]{figures/distribution_type.pdf}\label{fig:types}}
%   \caption{Distribution of malicious behaviors.}
%   \label{fig:distribution}
%    \vspace{1em}
% \end{figure}

\noindent \textbf{Malicious Behaviors Identified.}
Following the filtering of unsafe APIs/Libs, we perform extensive malicious behavior detection on these suspicious code snippets. 
So far, based on a three-month continuous detection on the Ant Group mirrored Hugging Face instance, \toolname{} has identified 91 malicious models and 9 malicious dataset loading scripts. Among the 91 malicious models, we found 76 Pickle variants and 15 models using Keras custom Lambda layers for malicious purposes. The publication dates of these malicious artifacts range from March 2022 to June 2024. \autoref{fig:monthly_malicious_behaviors} presents a classification of malicious behaviors based on code snippets extracted from these identified malicious artifacts, categorized through static analysis techniques and meticulous manual reviews by experienced researchers. The classification prominently includes remote control, sensitive information theft, and proof-of-concept. 
In distinguishing between proof-of-concept and actual malicious behaviors, we rely on detailed manual reviews. This process reveals that some codes initially flagged as malicious are, in fact, proof-of-concept experiments by researchers, posing no direct harm.
The statistics reveal a fluctuating but generally increasing trend in malicious behaviors over the observed period. We observe a significant increase in the latter half of the study period, with Q1 2024 and Q2 2024 showing the highest percentages of malicious artifacts. This trend suggests an escalating sophistication or frequency of malicious activities in recent months.

\subsection{Comparison with SOTA Techniques}
% \noindent \textbf{Qualitative Analysis.}
To contextualize the capabilities of \toolname{}, we conduct a qualitative comparison~(See \autoref{tab:qualitative}) with other SOTA techniques in PTM code poisoning detection. Existing tools like \textsc{Pickle Scanning}\cite{huggingface2024picklescanning}, \textsc{PickleScan}\cite{mmaitre2024picklescan}, and \textsc{Fickling}~\cite{trailofbits2021fickling} primarily focus on detecting unsafe libraries and API calls in pickle files. \textsc{Bhakti}\cite{dropbox2024bhakti} and \textsc{ModelScan}~\cite{protectai2023modelscan} extend to unsafe Lambda layer detection of TensorFlow and Keras models, but still concentrate on library and API-level analysis and fail to analyze dataset loading scripts.
In contrast, \toolname{} offers several distinctive features that set it apart from existing solutions. Unlike other tools that focus solely on unsafe libraries and API calls, \toolname{} performs analysis at the semantic level, allowing for a more nuanced and comprehensive detection of potential security threats. Moreover, \toolname{} is the only tool in our comparison that extends its analysis to dataset loading scripts, addressing a critical gap in the current security landscape of model hub ecosystems. Similar to \textsc{ModelScan}, \toolname{} supports various pickle variants as well as TensorFlow and Keras formats, enabling comprehensive security analysis across different model types.

% \noindent \textbf{Quantitative Analysis.}

\subsection{Case Studies}

\begin{figure}[t]
    \centering
    \includegraphics[width=\columnwidth]{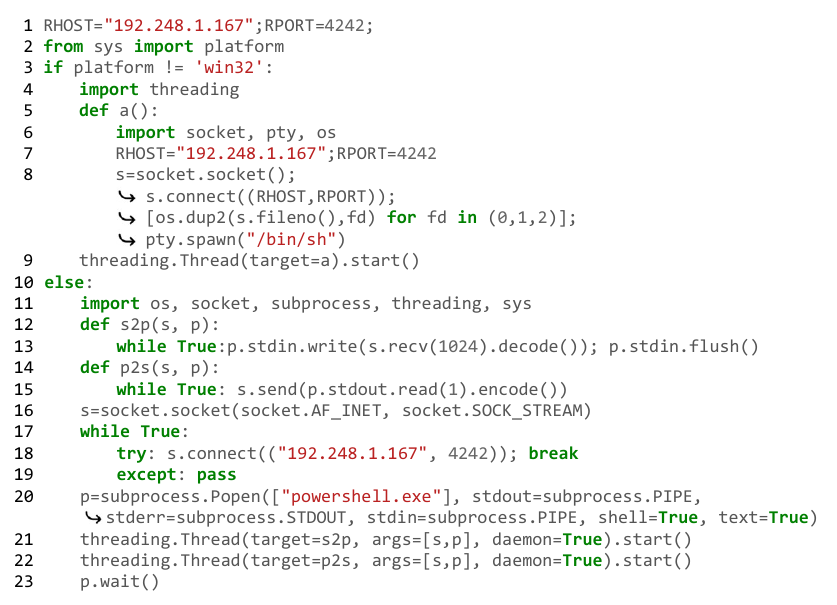}
    \caption{Code snippet injected into ``\texttt{star23/baller10}'', which establishes a reverse shell, enabling remote control.}
    \label{fig:case1-remote-control}
\end{figure}

\noindent \textbf{Case\#1:~Remote Control.}
As shown in \autoref{fig:case1-remote-control}, malicious code exists in a PyTorch model repository named ``\texttt{baller10}'', which establishes a reverse shell when the model is loaded, executing commands based on the operating system~(Windows or UNIX-like). The script first defines the attacker's host and port (line 1), then determines the operating system (lines 2-3). For non-Windows systems (lines 4-9), it creates a socket connection, redirects I/O, and spawns a shell. For Windows (lines 11-23), it establishes a connection to the attacker's machine and creates a PowerShell process with bidirectional communication. The malicious payload resembles those found in the previously identified ``\texttt{baller423/goober2}'' repository by JFrog~\cite{jfrog2024malmodel}, revealing a pattern of malicious code reuse and adaptation. Despite the subsequent deletion of the ``\texttt{baller423}'' account, the similarity in model name ``\texttt{baller10}'' suggests a possible connection. Notably, for the 10 malicious models created by ``\texttt{star23}'', our analysis unveils a broader attack strategy: these models' reverse shell commands point to different geographical locations, including Sri Lanka, Germany, and Poland, indicating that the attackers might use proxy servers to hide their real location. Despite being labeled ``for research use'' with warnings against downloading, these models successfully connect to external servers, posing significant security risks. This case highlights the real-world consequences of such attacks on unsuspecting users and emphasizes the importance of robust security protocols in PTM reuse workflows.

\begin{figure}[t]
    \centering
    \includegraphics[width=\columnwidth]{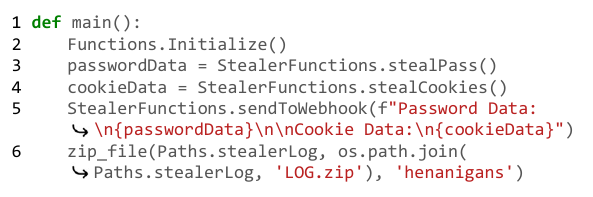}
    \caption{Dataset loading script in ``\texttt{Besthpz/best}'', which steals Chrome credentials and sends them to a remote server.}
    \vspace{2pt}
    \label{fig:case2-sensitive}
\end{figure}

\noindent \textbf{Case\#2:~Chrome Credential Stealer.}
This case examines a sophisticated malware newly discovered in the ``\texttt{Besthpz/best}'' repository, designed to steal credentials from Google Chrome browsers. The malware's main function~(See \autoref{fig:case2-sensitive}) executes a series of operations to extract and exfiltrate sensitive user data. Initially, it calls \texttt{Functions.Initialize}~(line 2) to prepare the environment, terminating any running Chrome processes and setting up necessary directories. The malware then proceeds to steal passwords and cookies using \texttt{StealerFunctions.stealPass}~(line 3) and \texttt{StealerFunctions.stealCookies}~(line 4) respectively. These functions decrypt and extract login credentials and cookie data from Chrome's local storage. The stolen information is then sent to a remote server using \texttt{StealerFunctions.sendToWebhook}~(line 5), potentially compromising user privacy and security. Finally, the malware creates a password-protected ZIP file containing the stolen data~(line 6), further obfuscating its activities.

\begin{figure}[ht]
    \centering
    \includegraphics[width=\columnwidth]{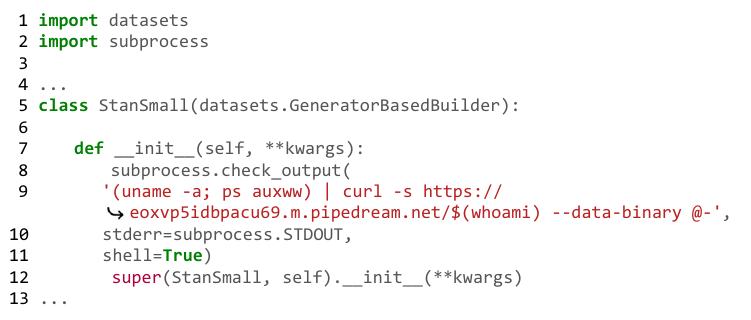}
    \caption{Dataset loading script in ``\texttt{Yash2998db/stan\_small}'', which leaks sensitive system information.}
    \label{fig:case3-sensitive}
\end{figure}

\noindent \textbf{Case\#3: Operating System Reconnaissance.}
Case\#3 a malicious loading script newly discovered in the ``\texttt{Yash2998db/stan\_small}'' dataset repository. The script contains suspicious code within its initialization method~(See~\autoref{fig:case3-sensitive}). Specifically, in the~\texttt{\_\_init\_\_} method of the \texttt{StanSmall} class~(line 7), the script executes a subprocess that collects and exfiltrates sensitive system information~(lines 8-9), which uses~\texttt{subprocess.check\_output} to run shell commands that gather system details~(\texttt{uname -a}) and information about running processes~(\texttt{ps auxww}). The collected system information is then sent to a remote server~(\texttt{eoxvp5idbpacu69.m.pipedream.net}) via a curl command, with the current user's identity~(whoami) appended to the URL.

\section{Discussion}
\noindent \textbf{Mitigation.}
Mitigating code poisoning attacks on model hubs requires a comprehensive approach combining platform-level security and developer vigilance. While Hugging Face has implemented pickle import scanning, this measure alone is insufficient due to its inability to perform deep semantic analysis of potentially malicious code. As for malicious dataset loading scripts, Hugging Face plans to disable the automatic execution of dataset loading scripts by default in their next major release, requiring users to explicitly set ``{\texttt{trust\_remote\_code=True}}'' for script-dependent datasets~\cite{huggingface2024loadingscripts}. Additionally, Keras has addressed vulnerabilities related to Lambda layers in version 2.13~\cite{cert2024keras,keras2024downgrade}, enhancing the security of models using this feature. Despite these improvements, developers must remain vigilant, adopting safer practices such as using secure model formats and treating unknown pre-trained models with caution, adhering to the principle that ``Models Are Codes''. 

\noindent \textbf{Generalizability and Scalability.}
While our study primarily focuses on the Hugging Face platform, the insights gained and methodologies developed are broadly applicable to other model hubs. The identified code poisoning attack vectors and proposed mitigation strategies are relevant across various platforms and frameworks. Our approach demonstrates the potential for large-scale analysis of models and datasets.

\noindent \textbf{Limitations}
While our study provides valuable insights into code poisoning attacks on model hubs, several limitations warrant consideration. Firstly, due to access permission restrictions, our analysis could not encompass all models and datasets on the platform, potentially leading to undetected malicious instances. Secondly, the collection of unsafe libraries and APIs, though informed by existing work like Pysa~\cite{facebook2024pysarules}, may not exhaustively cover all potential malicious exploits in the wild. Thirdly, although we have not encountered examples of obfuscation techniques used to evade static analysis in models, the possibility of such anti-analysis methods cannot be dismissed, drawing parallels from research on package manager poisoning~\cite{duan2021measuring,li2023malwukong}. Finally, we identify potentially malicious TensorFlow models by flagging those using unsafe operators, which may result in false positives. These limitations underscore the need for continuous refinement of detection methodologies and highlight the challenges in securing pre-trained model hubs against evolving threats.

\section{Related Work}
\noindent \textbf{Malicious Code Poisoning Attacks.}
Code poisoning attacks have been a persistent threat in software supply chains. Recent studies have explored these attacks in various contexts, including package managers~\cite{ohm2020backstabber,duan2021measuring,ladisa2023attacksok,huang2024donapi} and pre-trained model pipelines~\cite{zhang2021trojaning,hua2024malmodel,li2021deeppayload}. Ladisa et al.~\cite{ladisa2023attacksok} proposed a comprehensive taxonomy of attacks on open-source supply chains, covering 107 unique vectors linked to 94 real-world incidents. In the PTM domain, Hua et al.~\cite{hua2024malmodel} demonstrated how malicious payloads could be hidden in mobile deep learning models using black-box backdoor attacks. Building upon these studies, our work extends the current understanding by conducting the first systematic investigation of malicious code poisoning attacks specifically targeting pre-trained model hubs.

% \noindent \textbf{Insecure Deserialization Vulnerability.}
% Insecure deserialization has been recognized as a critical vulnerability in various software systems, with extensive research conducted in languages like Java~\cite{sayar2023rce,chen2024efficient,srivastava2023crystallizer}. 
% Our study builds on these findings, focusing on the implications of insecure deserialization across various pre-trained model formats.

\noindent \textbf{Security of Model Hubs.}
As model hubs have gained prominence, their security has become a growing concern. Zhou~\cite{blackhat2024pickle} examined insecure deserialization in pre-trained large model hubs, revealing risks in unsafe \texttt{pickle.loads} operations. Walker and Wood~\cite{blackhat2024confused} analyzed machine learning supply chain attacks, highlighting the danger of maliciously crafted model files. Jiang et al.~\cite{jiang2022security} studied artifacts and security features across multiple model hubs, exposing insufficient defenses for pre-trained models (PTMs). In a separate study, Jiang et al.~\cite{jiang2023exploring} investigated PTM naming practices on Hugging Face, introducing \textsc{DARA} for detecting naming anomalies.
Our work extends beyond these studies by providing the first systematic investigation of malicious code injection attacks specifically targeting pre-trained model hubs. We not only analyze vulnerabilities and attack vectors but also implement a detection pipeline deployed in a real-world industrial setting.
\section{Conclusion}
This paper presents the first systematic study of malicious code poisoning attacks on pre-trained model hubs, focusing on the Hugging Face. We developed \toolname{}, an end-to-end pipeline that addresses the limitations of existing tools through comprehensive analysis techniques. The deployment within Ant Group demonstrated its effectiveness in real-world industrial settings, uncovering 91 malicious models and 9 malicious dataset loading scripts among over 705K models and 176K datasets. These findings reveal significant security threats, including reverse shell attacks, credential theft, and system reconnaissance. Our work advances our understanding of vulnerabilities in the PTM supply chain and provides a practical solution for enhancing model hub security. 
\section*{Acknowledgment}
This work was supported by the National NSF of China (grants No.62072046), the Key R\&D Program of Hubei Province~(2023BAB017, 2023BAB079), the Knowledge Innovation Program of Wuhan-Basic Research (2022010801010083), Xiaomi Young Talents Program, and the research funding from MYbank (Ant Group).

\newpage

\balance
\bibliographystyle{ACM-Reference-Format}
\bibliography{main}

\end{document}